\documentclass{llncs}

\usepackage{url}
\usepackage{listings}
\usepackage{color}
\usepackage{setspace}
\usepackage{float}
\usepackage{subfigure}
\usepackage[fleqn]{amsmath}
\usepackage[utf8x]{inputenc}
\usepackage{multirow}
\usepackage{floatflt}
\usepackage{float}
\usepackage{tabularx}
\usepackage{algpseudocode}
\usepackage{setspace}
\usepackage{mathtools, cuted}
\usepackage{algpseudocode}
\usepackage{setspace}
\usepackage{amssymb}
\usepackage{balance}
\usepackage{graphics}
\usepackage{adjustbox}
\usepackage{mathtools, cuted}
\usepackage{caption}
\usepackage{url}

\definecolor{commentblue}{rgb}{0,0,1}

\begin{document}
\setstretch{0.97}
\title{PoliteCamera: Respecting Strangers' Privacy in Mobile Photographing}

\author{$^\dag$Ang Li, $^\S$Wei Du\thanks{This work was done when Wei Du was at the University of Arkansas.}, $^\dag$Qinghua Li}
\institute{$^\dag$Department of Computer Science and Computer Engineering, University of Arkansas\\$^\S$Department of Electrical and Computer Engineering, Michigan State University\\
	Email: $^\dag$\{angli,  qinghual\}@uark.edu, $^\S$duwei1@msu.edu}

\maketitle

\begin{abstract}
Camera is a standard on-board sensor of modern mobile phones. It makes photo taking popular due to its convenience and high resolution. However, when users take a photo of a scenery, a building or a target person, a stranger may also be unintentionally captured in the photo. Such photos expose the location and activity of strangers, and hence may breach their privacy. In this paper, we propose a cooperative mobile photographing scheme called PoliteCamera to protect strangers’ privacy. Through the cooperation between a photographer and a stranger, the stranger's face in a photo can be automatically blurred upon his request when the photo is taken. Since multiple strangers nearby the photographer might send out blurring requests but not all of them are in the photo, an adapted balanced convolutional neural network (ABCNN) is proposed to determine whether the requesting stranger is in the photo based on facial attributes. Evaluations demonstrate that the ABCNN can accurately predict facial attributes and PoliteCamera can provide accurate privacy protection for strangers.
\end{abstract}


\section{Introduction}\label{sec:intro}
\begin{figure}[h]
	\centering
	\subfigure[A stranger is included when the photographer pictures a building.]{
		\centering
		\includegraphics[scale=0.04]{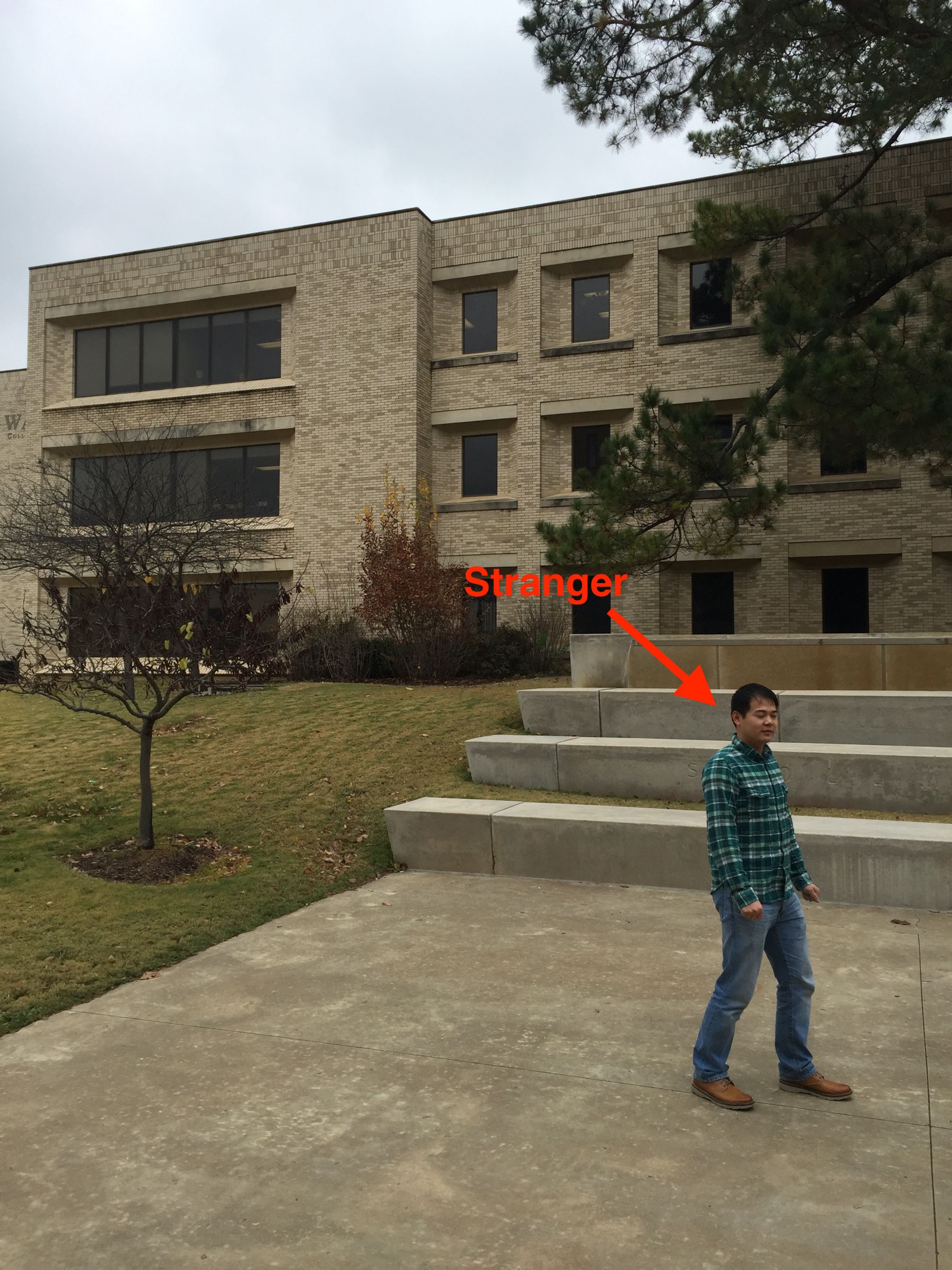}
		\label{fig:issue1}}
	\subfigure[Two strangers are included when the photographer pictures a target person.]{
		\centering
		\includegraphics[scale=0.04]{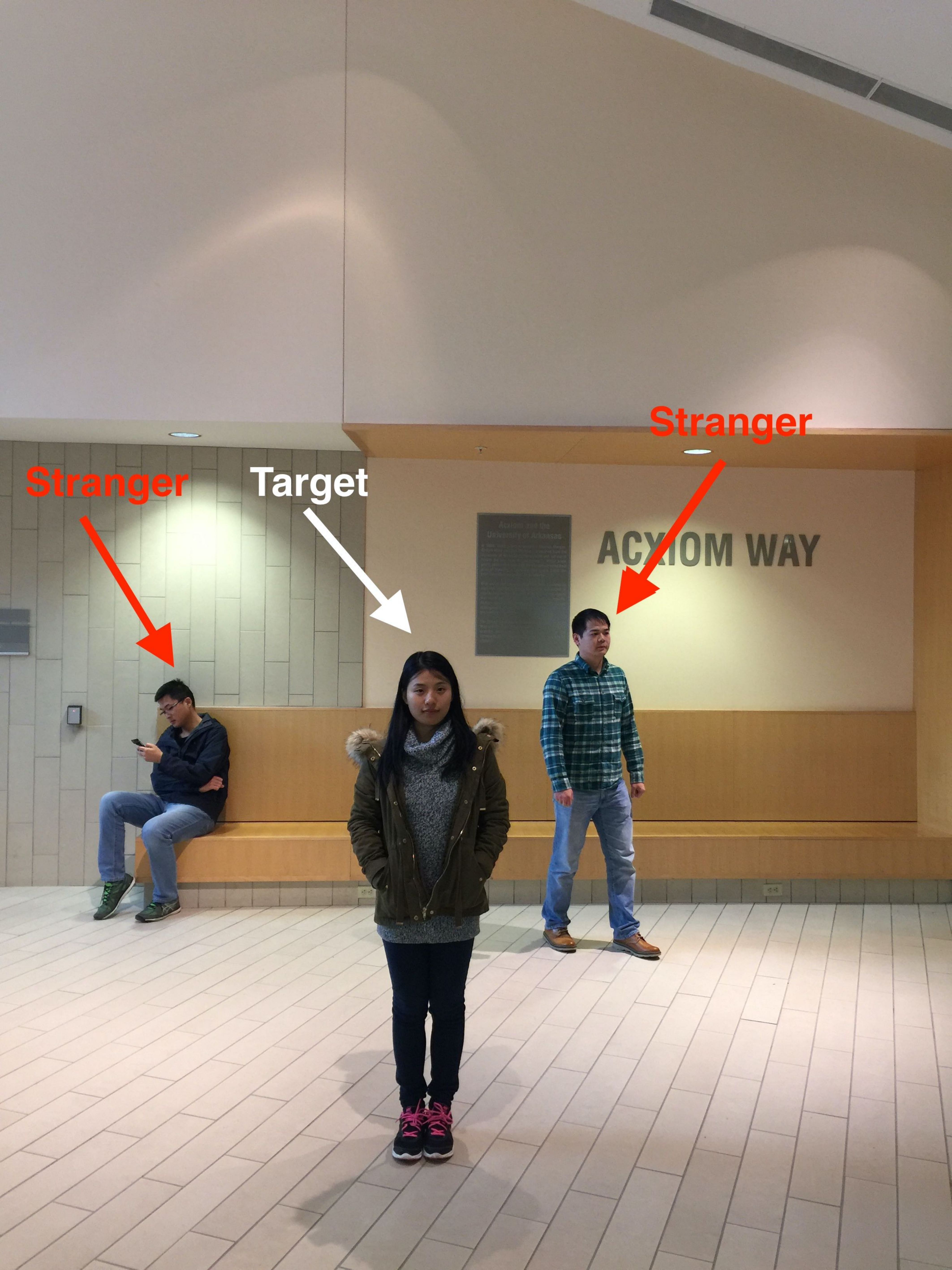}
		\label{fig:issue2}}
	\caption{Privacy issues with photos taken by mobile phones.}\label{fig:issues}\vspace{-0.15in}
\end{figure}

\vspace{-0.1in}
Nowadays mobile phones usually have built-in cameras that facilitate capturing photos. For instance, iPhone 7 is embedded with a 12-megapixel camera \cite{iphone7}. However, an increasing privacy concern has arisen as more and more pictures are taken in people's daily lives. When a user takes a photo of a scenery or a friend with a mobile phone, it is likely that a stranger can also be accidentally included in the photo, with the face clearly recognizable. Fig. \ref{fig:issues} illustrates two examples. In Fig. \ref{fig:issue1}, the building is the target but a stranger is captured; in Fig. \ref{fig:issue2}, the photographer intends to picture the target person but two strangers are accidentally included. In these examples, the photo can breach the stranger's privacy by revealing the stranger's location and activity. Thus strangers’ privacy should be protected.
	
Based on advanced techniques in computer vision, there exist several applications which can blur faces in a photo, such as ObscuraCam \cite{obscuracam}, Point Blur \cite{pointblur} and \cite{blurimage}. However, none of these commercial applications can inform the inclusion of a stranger in a photo and allow him to decide whether to blur his face or not. Only the photographer is allowed to determine the necessity of blurring the stranger's face.

Several recent works have been done to protect strangers’ privacy in photos through blurring their faces. They differ in the way of determining whether a stranger is in the photo or not. Our previous work \cite{li2016privacycamera} checks whether a stranger is in a photo or not based on GPS locations of the photographer and the stranger. Due to the dependence on GPS location, it does not work well indoor due to the unavailability or inaccuracy of GPS. He et al. \cite{wang2015visually} design a system for protecting photo privacy that identifies a stranger in a photo by recognizing his motion patterns and visual appearance (e.g., clothes color) profiled into the system in advance. However, users' visual fingerprints need to be updated whenever they change (e.g., changing clothes), which is not convenient. Zhang et al. \cite{zhang2016privacy} propose a server-based system to protect privacy of photographed users that compares the portrait of a user uploaded to the server and the portrait of the persons included in photos. Their scheme considers full portrait captured in the photo (i.e., the whole body), which is quite different from this paper that only considers face. Also, their scheme assumes a trusted server from the privacy perspective, which is not always available. 
	
In this paper, we use facial attributes that do not change frequently (e.g., black hair or blond hair) to determine whether a stranger is in a photo or not. Since such facial attributes are relatively stable, if a person is in a photo, by comparing the faces in the photo with his recent profile photo in facial attributes, the person can be correctly matched to his face in the photo. Also, in photographing scenarios, it is not very likely that the facial attributes of two nearby strangers are exactly same, since the number of persons in a limited geographic area around the photographer is usually not large. That means if facial attributes can be accurately identified from photos, mismatch between faces and strangers will be of a low chance. Thus intuitively facial attribute-based face-stranger matching is a promising method to explore. 

Based on facial attributes, we design a cooperative scheme \textit{PoliteCamera} to protect the privacy of strangers who are unintentionally included in photos taken by mobile phones. PoliteCamera works as an application on the mobile phones for both the photographer and the stranger. When a photographer takes a photo, he (via the mobile phone) will notify nearby strangers of the potential risk of being included in the photo via peer-to-peer short-range wireless communications (e.g., WiFi Direct \cite{wifidirect}). If a stranger prefers not to be included in the photo, he can send a blurring request to the photographer together with his facial attributes included in the request. The photographer will check whether the requesting stranger’s face appears in the photo or not based on the facial attributes sent from the stranger and the facial attributes of faces captured in the photo. If the attributes of a face in the photo match those of the requesting stranger, that face is considered to be the stranger's and it will be blurred in the photo. 

The set of facial attributes will be carefully selected so that a combination of attribute values is specific enough to differentiate different strangers nearby the photographer but is not specific enough to uniquely identify who the requesting stranger is in the real world. The number of possible attribute value combinations should be reasonably large (e.g., tens of thousand). Then the probability for two different strangers to have the same combination is low, since the number of strangers around a photographer is usually small. The number of possible combinations should also not be too large. In this way, each combination could be owned by many people in the real world, and thus cannot be used to infer who the stranger is. As described later, approximate match instead of exact match will be used in PoliteCamera, which makes linking multiple appearances of the same person difficult. Thus, the privacy risk of re-identification will be low. Moreover, privacy-preserving computing technologies can also be applied to complete the matching of facial attributes without sending the stranger's facial attributes to the photographer in cleartext, and in this way further protect the stranger's facial attributes from the photographer (see Section \ref{subsec:design_determination} for a discussion). 

The privacy protection offered by PoliteCamera is based on the cooperation between photographers and strangers. Although these two roles are separately discussed, real-world users can take either role in different scenarios. Since every user can be a stranger in many scenarios, users have a motivation to use this system, and participation in this system means mutually protecting each other’s privacy and benefiting everyone including self. This inter-user cooperation design is also motivated by many real-world systems such as collaborative filtering recommender systems \cite{schafer2007collaborative} and peer-to-peer video streaming systems \cite{liu2008survey}. Users' privacy can be better protected when more people use this system. Although it is not a perfect solution for the problem, it still significantly advances the state of the art in this domain. 


The contribution of this paper is summarized as follows:
\begin{itemize}
	\item We propose a facial attribute-based system PoliteCamera for protecting strangers' privacy in mobile photographing. To the best of our knowledge, PoliteCamera is the first scheme that makes nearby strangers aware of possible inclusion in a photo when the photo is being taken, allowing them to determine whether to blur their face in the photo or not, and protects strangers' privacy under both indoor and outdoor scenarios, without using any trusted server, human gesture, or special wearables.
	\item We design a novel adapted balanced convolutional neural network (ABCNN) that can simultaneously predict multiple facial attributes from a photo, and use it to determine the existence of requesting strangers in a photo.
	\item To avoid identifying the real target persons of a photo as a stranger, a heuristic approach is employed to effectively filter targets to prevent incorrect blurring. 
	\item The proposed system is implemented, and extensively evaluated on real datasets and in the field. Experimental results show the excellent performance of the system. 
\end{itemize}

The rest of the paper is organized as follows.  Section \ref{sec:systemdesign} introduces the design of PoliteCamera. Section \ref{sec:implementation} presents implementation. Section \ref{sec:evaluations} shows evaluation results. Section \ref{sec:related} reviews related work. Section \ref{sec:conclusion} concludes the paper.\vspace{-0.15in}

\section{System Design}\label{sec:systemdesign}\vspace{-0.1in}
This section describes the design of PoliteCamera.\vspace{-0.15in}

\subsection{System Overview}\label{subsec:overview}\vspace{-0.1in}
Three types of entities are involved in the system: the \textit{photographer} who takes a photo, the \textit{target} who is intentionally captured by the photographer, and the \textit{stranger} who is near the target and might be accidentally included in the photo. 

The system is designed to protect the stranger's privacy by giving an option to the stranger to opt out from the photo. The general idea is that the system notifies nearby strangers the possible inclusion in a photo, and blurs a stranger's face if the stranger sends a blurring request. A naive approach is to blur every stranger's face in the photo. However, this is not an ideal solution, since blurring will inevitably affect the quality of the photo. To minimize the effect on photo quality, our design only blurs a stranger's face if he requests to do so. We assume PoliteCamera is installed on both the photographer’s and the stranger's mobile phone. Each user of PoliteCamera provides one of his photos to the PoliteCamera app upon the installation of the system. Each user’s facial attributes are learned from this base photo and stored in the system for future use. (The base photo can be updated by the user but this does not need to be done frequently since facial attributes do not change frequently.) When a stranger requests a photographer to blur his face, he can send these attributes to the photographer and the photographer will determine whether his face is in the photo based on these facial attributes and blur his face if so.


There are two challenges with the approach. Firstly, there might be multiple nearby strangers who receive the notification of potential privacy leakage by the photo. Some of them may request to blur their faces but others may not request so. Hence, we need to determine if the requesting stranger’s face is in the photo or not, which is not trivial. Secondly, when the target is a single person or multiple persons, we need to keep the target unblurred even if the target's phone mistakenly sends out a blurring request. Telling the target from the stranger is necessary but difficult.


\subsection{The Architecture and Workflow of PoliteCamera}
\begin{figure}
	\centering
	\includegraphics[scale=0.25]{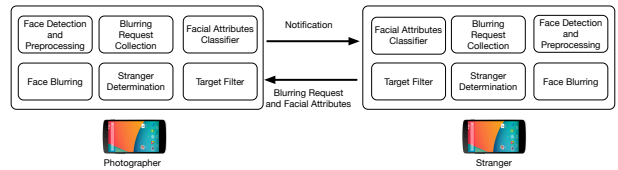}
	\caption{The architecture of PoliteCamera.}\vspace{-0.15in}
	\label{fig:overview}
\end{figure}

\vspace{-0.1in}
As Fig. \ref{fig:overview} shows, the system consists of six major modules: \textit{face detection and preprocessing}, \textit{blurring request and collection}, \textit{facial attributes classifier}, \textit{target filter}, \textit{stranger determination} and \textit{face blurring}. When a photographer takes a photo, the face detection module will run on the captured image. If any face is detected, the notification of possible inclusion in the photo will be sent to nearby strangers via peer-to-peer short-range wireless communications. If a stranger would like to blur his face in the photo, he sends a blurring request to the photographer. To help the photographer determine if the requesting stranger is in the photo, this stranger also sends his pre-computed facial attributes (e.g., gender, obtained from his face image when initializing the PoliteCamera app). Upon receiving blurring requests, the photographer crops all the faces in the picture, and then feed them into the pre-trained facial attributes classifier. By comparing the facial attributes of requesting strangers and the attributes of detected faces in the photo, the stranger determination module of photographer can identify those requesting strangers captured in the photo. If a requesting stranger is in the photo, the face blurring module of the photographer smoothly blurs the corresponding face; otherwise, the request is ignored. In case the target mistakenly sends a blurring request, the target filter module distinguishes the target from the stranger based on specific defined rules, and keeps the target unblurred in the photo.

The design of PoliteCamera depends on several available technologies in mobile phones. In particular, face detection and preprocessing can be implemented using APIs provided by the operating system on mobile phones, such as the \textit{FaceDetector} APIs in Android SDK. Similarly, peer-to-peer short-range wireless communications can be set up by available technologies on most modern mobile phones, such as  WiFi Direct \cite{wifidirect} and Bluetooth. We will introduce the implementation of these two modules in Section \ref{sec:implementation}. Next, we will illustrate more details about the rest four modules.\vspace{-0.1in}

\subsection{Facial Attribute Classifier}\label{subsec:desgin_classifier}\vspace{-0.1in}
Given an input face image in pre-defined dimensions, this module aims to  simultaneously output a set of  facial attributes associated with this input image. In particular, each facial attribute is a binary label, where +1 indicates the presence of the corresponding attribute, and -1 means its absence. In this paper, we propose to train a facial attribute classifier through the ABCNN model where a weighted objective function is constructed to maximize the prediction accuracy.

Formally, let $\mathbb{I}$ be the set of input images, and $N$ be the number of facial attributes. For a given image $x \in \mathbb{I}$, let $y_i \in \{-1,+1\}$ be the binary label of the $i$th attribute, where $i \in \{1,2,\ldots,N\}$ is the index of facial attributes. Let $\mathbb{H}$ be the hypothesis space of possible decision functions, and $f_i(\theta^{T}x)$ be the decision function, where $\theta=\{\theta_1,\theta_2,\ldots,\theta_N\}$ is the network weights. Hence, the loss function of the $i$th facial attribute can be defined as $L_i(f_i(\theta^{T}x),y_i)$. Let $\mathbb{E}(L_i)$ be the expected loss over the range of inputs $\mathbb{I}$. Then the optimization task is to minimize the expected squared error for each attribute.

\begin{equation}\label{equ1}
\forall i: f_i=\underset{f_i\in \mathbb{H}}{\arg\min} \hspace{0.1cm}\mathbb{E}(L_i)\vspace{-0.15in}
\end{equation}
For each input $x$ and attribute $i$, the corresponding classification result $c_i(x)$ and the according accuracy $acc_i(x)$ can be obtained from the output of $f_i(x)$ described as:

\begin{equation}\label{equ2}
	\nonumber
		c_i(x)= 
		\begin{cases}
		+1 & f_i(x)>0\\
		-1              & \text{otherwise,}
		\end{cases}\quad \vspace{-0.15in}
\end{equation}
		and
\begin{equation}
		acc_i(x,y)=
		\begin{cases}
		+1 & y_i(x)c_i(x)>0\\
		0 & \text{otherwise}
		\end{cases}\vspace{-0.15in}
\end{equation}

As discussed above, the traditional approach treats facial attributes as $ N $ independent tasks, and each classifier is trained independently. The typical loss function for the $ i $th facial attribute is constructed by choosing the hinge-loss function, which is shown as:

\begin{equation}\label{equ3}
\resizebox{.9\hsize}{!}{$
\underset{\theta_i}{\arg\min}L_i(f_i(\theta^{T}x),y_i)=\underset{\theta_i}{\arg\min}(\max(0,1-y_i(x)f_i(\theta^{T}x)))$}
\end{equation}



However, a problem with the traditional approach is that training independent classifiers cannot learn the latent correlations between attributes. To exploit such correlations, the classifier should be constructed to learn all of these facial attributes simultaneously. In addition, the attribute label distribution in the training set should match with the corresponding distribution in the testing set. Therefore, it is necessary to balance the dataset to train a better classifier. One way to obtain a balanced dataset is to perfectly collect evenly distributed dataset of images for each attribute. However, it will cause extra efforts since most of data in real application is not evenly distributed, and finding such dataset may be very challenging especially at a large scale. An alternative solution is to modify the loss function in order to simulate a balanced dataset. In our proposed ABCNN, some changes are made to the objective function to address the imbalance between the training dataset and the test dataset. Specifically, a mixed objective function is proposed by considering the distribution difference between training data and testing data as adapted weights. Firstly, the training distribution $S_i$ for each attribute $i$ is computed by calculating the fraction of positive samples $Train_i^+$ ($0<Train_i^+<1$) and fraction of negative samples $Train_i^-$ ($0<Train_i^-<1$)  in the training set. Given the binary testing target distribution $Target_i^+$ and $Target_i^-$ (where $Target_i^+ + Target_i^-=1$), an adapted weight is assigned for each class of attribute $i$, as shown in Eq. (4) and Eq. (5):
\begin{equation}\label{equ4}
p(i|+1) = 1 + \dfrac{\Delta T^+}{Target_i^+ + Train_i^+} 
\end{equation} 
\begin{equation}
p(i|-1) = 1 + \dfrac{\Delta T^-}{Target_i^- + Train_i^-} 
\end{equation}
where $\Delta T^+ = Target_i^+ - Train_i^+$ and $\Delta T^- = Target_i^- - Train_i^-$. It can be seen from the above equations that we will increase the weight of the $ i $th facial attribute if the fraction of positive or negative labels in the training data is less than the testing data. The intuition is that the increment of those weights will help balance the distribution difference between training data and testing data. Correspondingly, we will decrease the fraction weights of positive or negative labels in the training data if it is higher than that in the testing data. Then, these adapted weights are incorporated into the mixed objective function. Instead of using the hinge-loss function, a weighted mixed task square error is adopted as the loss function, and the optimization problem of ABCNN can be expressed as:
\begin{equation}\label{equ5}
\resizebox{.9\hsize}{!}{$
\forall i:\underset{f_i\in \mathbb{H}}{\arg\min} \hspace{0.1cm}\mathbb{E}(L(x,y))=\underset{f_i\in \mathbb{H}}{\arg\min}\hspace{0.1cm}\mathbb{E}(\sum_{i=1}^{N}p(i|y_i(x))||f_i(x)-y_i(x)||^2)$}
\end{equation}
The optimization problem aims to find the optimal decision function $f$ that has the smallest error between predictions and target labels. Over an $M$-element training set X with labels Y, from Eq. (6) we can get:
\begin{equation}\label{equ6}
\resizebox{.9\hsize}{!}{
$\forall i:\underset{f_i\in \mathbb{H}}{\arg\min}\hspace{0.1cm}\mathbb{E}(L(X,Y))=\underset{f_i\in \mathbb{H}}{\arg\min}\hspace{0.1cm}\mathbb{E}(\sum_{j=1}^{M}\sum_{i=1}^{N}p(i|Y_ji(x))||f_i(X_j)-Y_ji||^2)$}
\end{equation}

The ABCNN architecture can be built by replacing the standard loss layer of a deep convolution neural network (DCNN) with a layer implementing Eq. (7). After the above classifier is trained, we can predict facial attributes by inputing a face image with fixed dimensions (which are consistent with that of training images) to the classifier.\vspace{-0.15in}

\subsection{Stranger Determination}\label{subsec:design_determination}\vspace{-0.1in}
This module aims to determine if a requesting stranger is included in the photo or not and which face matches the stranger. This is done though thresholding the difference between the facial attributes of the detected faces and those of the requesting stranger. In fact, facial attributes predicted by the classifier is a vector of binary values, where $+$1 indicates the presence of the corresponding attribute, while $-$1 represents its absence. The difference is defined as the number of different attributes between two faces under the same set of attributes. Formally, let $N$ be the number of attributes associated with a face. For a given face, its corresponding attributes vector $V=[a_1,\ldots,a_N]$, where $a_i\in\{-1,+1\}$ represents the $i$th facial attribute. We use $V_r$ and $V_s$ to represent the facial attributes of the requesting stranger and a specific detected face respectively. The inner product of $V_r$ and $V_s$ is $V_r\cdotp V_s=\sum_{i=1}^{N}V_r[i]V_s[i]$. If all the attributes are identical that inner product should be $N$. The $ V_r[i]V_s[i] $ is $-$1 only when the $i$th attribute in $V_r$ and the $i$th attribute in $V_s$ are different. Hence, the difference can be obtained as: 
\begin{equation}\label{equ7}
diff=\frac{N-V_r\cdotp V_s}{2}
\end{equation}
As discussed before the facial attributes cannot be used to uniquely identify a stranger. In order to further protect the stranger's facial attributes from the photographer, inner product computation can be done with a two-party privacy-preserving scheme \cite{dong2011secure}. Usually the predication results from two images from the same person cannot match exactly due to angle difference or some other reasons. Thus a threshold is set to tolerate such minor deviations. The rule is that only the difference between facial attributes of the requesting stranger and any specific detected face is less than or equal to the threshold, we consider the detected face belongs to the requesting stranger. Our evaluations show that it is a good choice to set 1 as the threshold. \vspace{-0.15in}

\subsection{Target Filter}\label{subsec:design_filter}\vspace{-0.1in}
This module is designed to distinguish the target from the stranger in a photo, so that the target's face will not be blurred even if the target mistakenly sends a blurring request. Specifically, if the target of a photo is one or multiple persons, the task is filtering out the targeted faces; if the target is a building or something else, we would like to avoid the stranger being mistakenly identified as the target. Therefore, a heuristic approach is proposed to achieve this goal. Based on our observations from real-world experience, the target is usually associated with the following properties in the photo:
\begin{itemize}
	\item One common goal of taking photos is recording beautiful moments. The target is likely to be smiling when he is being pictured, since smiles make a person more attractive and confident.	
	\item The photographer usually intentionally makes the target's face significantly larger than others who are accidentally included in the photo. For instance, if a stranger is too close to the camera and hence his face is larger than the target's, the photographer will usually stop picturing or move a little bit so that the target is better captured into the photo. Moreover, considering that there might be multiple targets appearing in the photo but with slightly different face sizes (e.g., a group of people taking a picture), we expect to filter all targets in the photo by comparing a detected face with the largest face in the photo, which is considered as one of the targets' faces by default. If the size difference is less than a pre-defined threshold, we consider the detected face as one of the target faces.
	\item Similarly, the photographer usually puts the target in a dominant position of the photo. The central region is one of the most popular options, which can highlight the target in the photo. 
\end{itemize}

Consequently, \textbf{smiling}, \textbf{face size} and \textbf{face position} can facilitate determining if a face belongs to the target or not. Based on these observations, we propose three rules to determine whether a person in the photo is a target or not.
\begin{enumerate}
	\item The person is smiling.
	\item The person's face is the largest one in the photo or slightly smaller than the largest one by a pre-defined threshold. Based on our test, we find that the average size difference between two targets' faces in a photo is around 10\%. Hence, if more than one face is detected, we compare the largest one with the others. If the size difference between the largest one and a certain face is less than or equal to 10\%, we consider that face as one target face. Otherwise, the detected face will not be treated as a target.
	\item The person's face appears at the central region of the photo. The central region is defined as the middle section of horizontal trisections of a photo.  
\end{enumerate}
However, it is too strict if we determine a detected face is the target only when all those three rules are satisfied, since sometimes not all of them are satisfied. For instance, the target is not always smiling when the photo is taken. Considering this, we determine that the face is the target if at least two of the three rules are satisfied. \vspace{-0.15in}

\subsection{Face Blurring}\label{subsec:design_faceblurring}\vspace{-0.1in}
The purpose of face blurring is to mask the features of a face in order to make the face not recognizable, without degrading the quality of photo much. Similar to our previous work \cite{li2016privacycamera}, we adopt an approach based on the Gaussian Blur algorithm \cite{gaussianblur1} to smoothly blur faces. To conduct face blurring, we need to determine a blurring area in the face enclosing the main identifiable features of the face. In particular, we draw a square whose side length is 2.4 times of the distance between eyes, and whose center is the middle point between eyes. Then the Guassian Blur operation can be performed in the square blurring area.\vspace{-0.15in}

\section{Implementation}\label{sec:implementation}\vspace{-0.1in}
The facial attribute classifier was implemented using Python 2.7 and MxNet \cite{chen2015mxnet}, which is an open-source deep learning framework. WiFi Direct was used to conduct peer-to-peer communications between the stranger and the photographer. The face blurring module was implemented as same as our previous work \cite{li2016privacycamera}, so some details are omitted here.

\subsection{Face Detection and Preprocessing}\label{subsec:implement_face}\vspace{-0.05in}
Face detection is based on the \textit{FaceDetector} class provided in Android SDK. Faces in an image can be detected by calling the \textit{findFaces} method of \textit{FaceDetector}. This method detects faces by finding pupils in the image, and returns a number of detected faces into an array of FaceDetector.Faces class. For each instance of \textit{Face} class, the distance between two eyes of a face and the coordinate of the middle point between two eyes can be obtained. Then we crop each detected face with a square area, which is the same as the blurring square described in Section \ref{subsec:design_faceblurring}. Also, the size of the cropped square is used to represent the size of the corresponding face in the target filter module. To prepare for target filtering, we need to detect the position of each face in the photo. To do so, we evenly divide the picture into three regions (left, middle, right) along the horizontal direction. Then for each detected face, we calculate the middle point between its eyes. If the middle point is located in the middle region, we say this face is in the central region.

\subsection{Facial Attribute Classifier}\label{subsec:implementation_classifier}\vspace{-0.05in}
This module aims to predict a set of facial attributes from a given face image. As described in Section \ref{subsec:desgin_classifier}, we use ABCNN to predict the facial attributes and ABCNN is implemented by the Python interface of MxNet \cite{chen2015mxnet}. In particular, we build the ABCNN network by replacing the final loss layer of a 16-layer VGG network from \cite{simonyan2014very} by the loss function in Eq. (7), and the architecture shown in Fig. 3. The architecture consists of 16 weight layers, including 13 convolution layers and 3 fully connected layers, which are associated with over one million weights. Since the network only accepts RGB image input with dimensions of 128*128 pixels, each cropped face obtained from the face detection and preprocessing module should be scaled to that size before being sent into this classifier. 

%
\begin{figure}[h]
		\centering
		\includegraphics[scale=0.3]{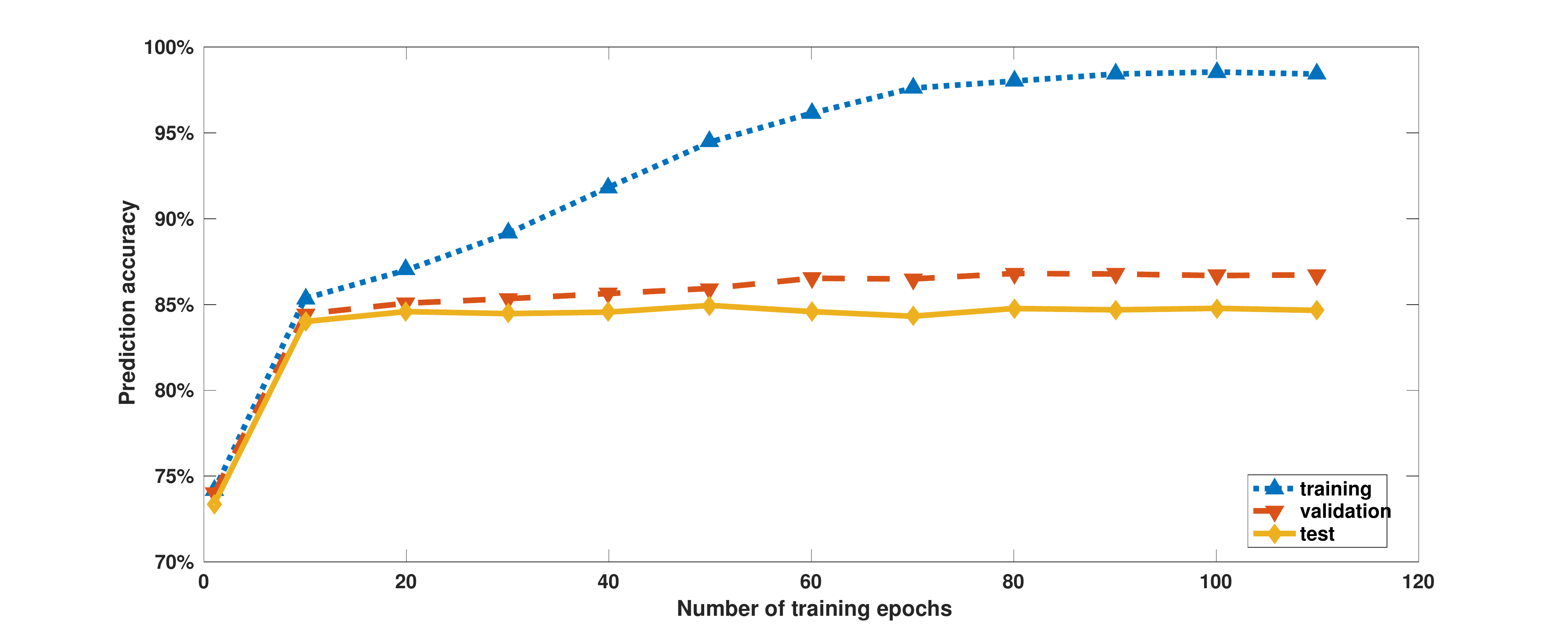}
		\caption{Architecture of the ABCNN network.}\vspace{-0.15in}
		\label{fig:moon}
\end{figure}

In this paper, the ABCNN network is trained on the CelebA dataset \cite{liu2015deep}, which is a large-scale facial attributes dataset. It contains 20 images for each of over 10K celebrities, hence with a total of more than 200K images. The first 160K images are used for training, and the remaining 40K images are used for validation and testing, specifically, 20K for validation and 20K for testing. For our implementation, we use a set of pre-cropped and aligned face images provided by the CelebA dataset, and scale the dimensions of training RGB images from 178*218 pixels to 128*128 pixels. Each image in the CelebA dataset is annotated with binary labels of 40 facial attributes (e.g., `Young' and `Male'). However, in this work, we choose 16 out of the 40 attributes that do not change frequently for the same person as our considered attributes. The 16 chosen facial attributes include \{\textit{Arched Eyebrows}, \textit{Bushy Eyebrows}, \textit{Big Lips}, \textit{Big Nose}, \textit{Point Nose}, \textit{Black Hair}, \textit{Blond Hair}, \textit{Brown Hair}, \textit{Gray Hair}, \textit{Eyeglasses}, \textit{Bald}, \textit{High Cheekbones}, \textit{Narrow Eyes}, \textit{Oval Face}, \textit{Male}, \textit{Young} \}. In addition, since the `smiling' attribute is required for target filtering, we also add it into the classifier (note that it is not used for stranger determination but only for target filtering).\vspace{-0.15in}

\section{Evaluations}\label{sec:evaluations}\vspace{-0.05in}
To train the classifier, we set the batch size to 384 images per training iteration, and hence the training process requires approximately 420 iterations to finish a full epoch on the training set. The learning rate is initialized as 0.05, and reduced by a factor of 0.8 every four epochs until it decays to 0.000001. We train the ABCNN for 110 epochs with all images from training set on two NVidia K80 GPUs.\vspace{-0.15in}

\subsection{Model Selection}\vspace{-0.05in}
Classification accuracy is defined as the number of correctly predicted cases divided by the number of testing images. From Eq. (2), we can derive the classification accuracy of each attribute $i$:
\begin{equation}\label{equ9}
e_i(X,Y)=\frac{1}{N_{test}}\sum_{j=1}^{N_{test}}acc_i(X_j,Y_j)
\end{equation}
Consequently, we can evaluate the average classification accuracy by calculating the average classification accuracy over all the $N$ attributes:
\begin{equation}\label{equ10}
E(X,Y)=\frac{1}{N}\sum_{i=1}^{N}e_i(X,Y)
\end{equation}

\begin{figure}[h]
	\centering
	\includegraphics[scale=0.25]{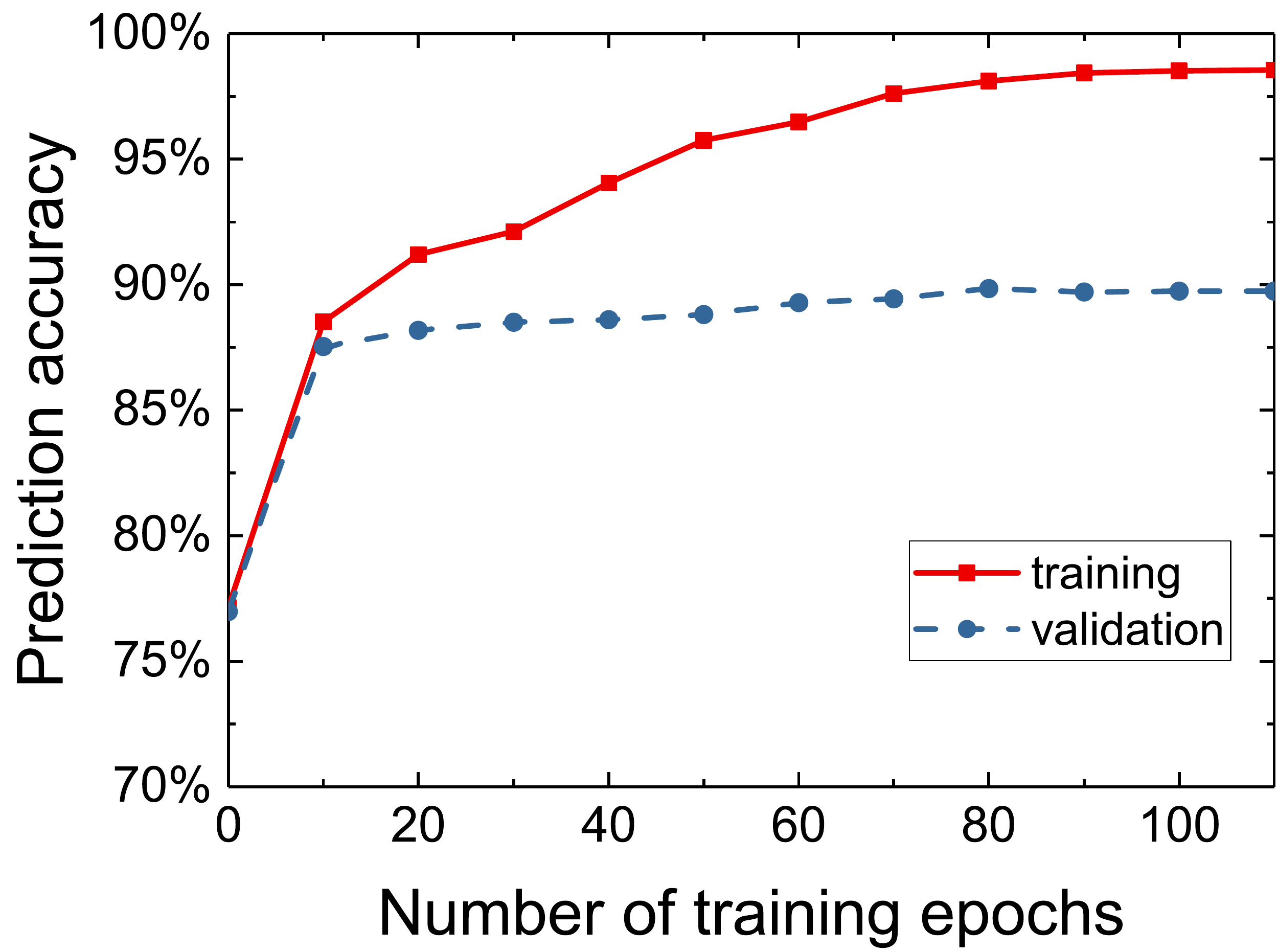}
	\caption{Average classification accuracy vs training epochs.}\vspace{-0.15in}
	\label{fig:accuray}
\end{figure}

The ABCNN prediction model is trained on the training dataset, but the number of training epochs needed is determined based on the validation dataset. Specifically, the accuracy trend when the number of training epochs increases is shown in Fig. \ref{fig:accuray}. As the training continues, the accuracy over the training dataset keeps increasing. However, training for more epochs means higher cost. Thus, based on the maximum accuracy over the validation dataset, we stop training the ABCNN network after 80 epochs (with $89.84\%$ validation accuracy) and use the resulted model for performance evaluations in order to guarantee the coverage of the model without too high cost. 

\begin{figure*}[h]
	\centering
	\includegraphics[scale=0.2]{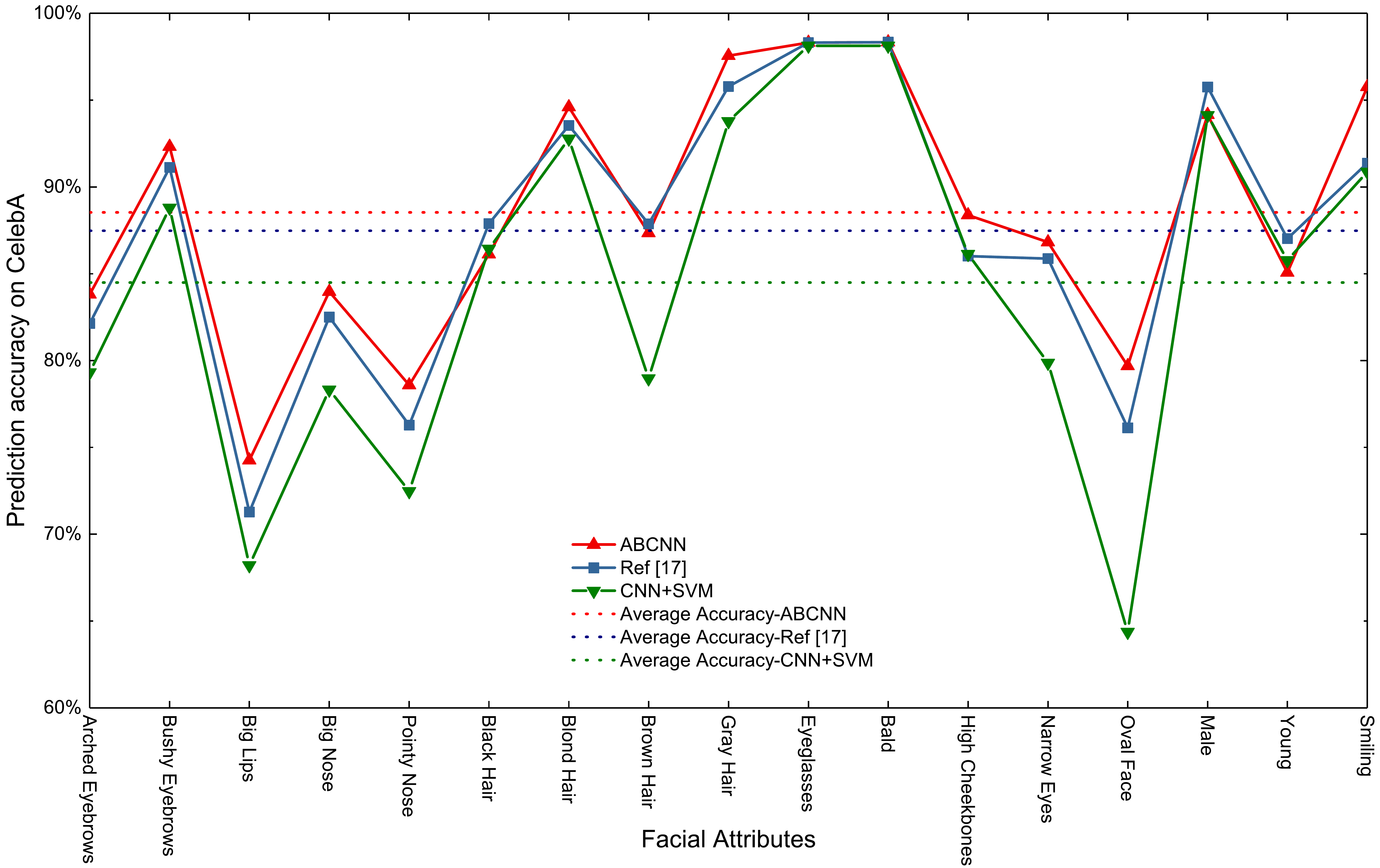}
	\caption{Classification accuracy of each attribute and average accuracy in testing.}\vspace{-0.15in}
	\label{fig:attribute_accuracy}
\end{figure*}

Then based on Eq. (9) we evaluate the classification accuracy of each facial attribute on the test dataset, including 16 attributes used for stranger determination and the `Smiling' attribute for target filtering. The average accuracy over those 16 attributes is also tested according to Eq. (10). As Fig. \ref{fig:attribute_accuracy} shows, the average accuracy is $88.53\%$ (see the horizontal dashed line) which is pretty high. Out of the first 16 facial attributes, 6 attributes outperform the average performance, including \textit{Bushy Eyebrows},  \textit{Black Hair}, \textit{Blond Hair},  \textit{Gray Hair}, \textit{Eyeglasses}, \textit{Bald} and  \textit{Male}. For example, the classification accuracies of \textit{Eyeglasses} and  \textit{Bald} achieves $98.31\%$ and $98.34\%$, respectively.

To measure the performance of our proposed ABCNN in predicting the facial attributes, we compared it with the state-of-art algorithm proposed in [17]. They also construct a multi-task training classifier and the corresponding facial attribute prediction and average accuracy are represented with the blue dashed line and the horizontal blue solid line in Fig. 5, respectively. In addition, we also compared the proposed ABCNN with \cite{liu2015deep} which uses the basic CNN model to select features and inputs them to the SVM classifier for training. Its performance is displayed by green line and green dashed line for facial attributes prediction accuracy and average accuracy, respectively. ABCNN outperforms both the multi-task training classifier in [17] and the CNN-SVM model [13].   

\subsection{Classification Consistency}
\begin{figure}[h]
	\centering
	\subfigure[First face image]{
		\centering
		\includegraphics[scale=0.35]{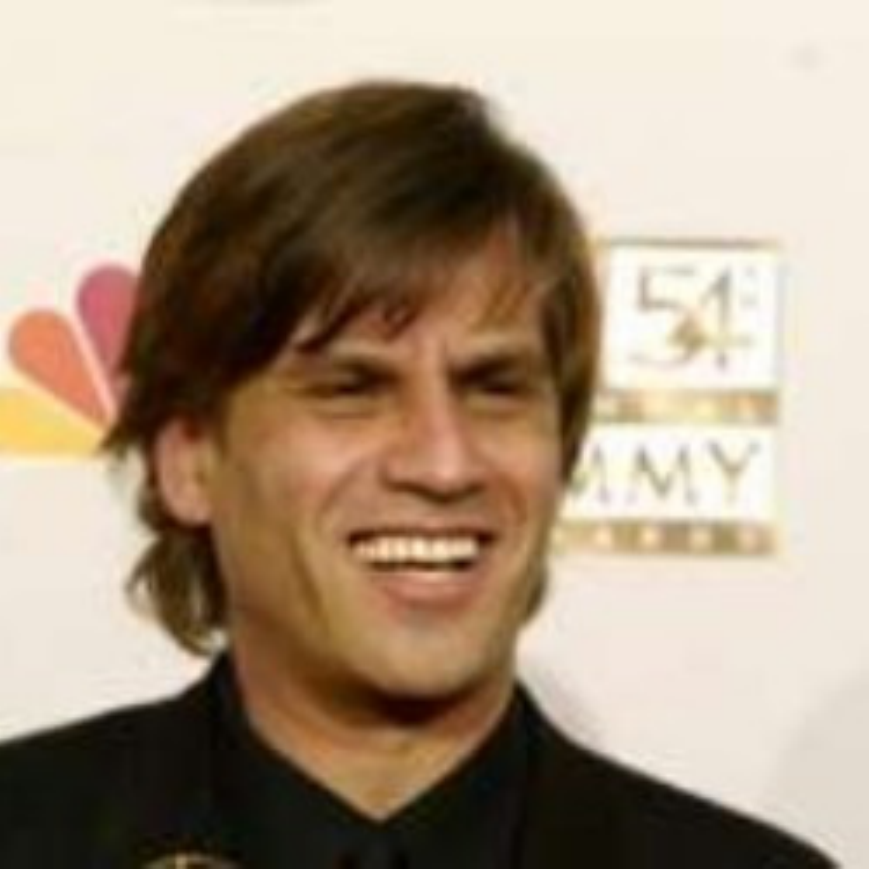}
		\label{fig:entity1}}
	\subfigure[Second face image]{
		\centering
		\includegraphics[scale=0.35]{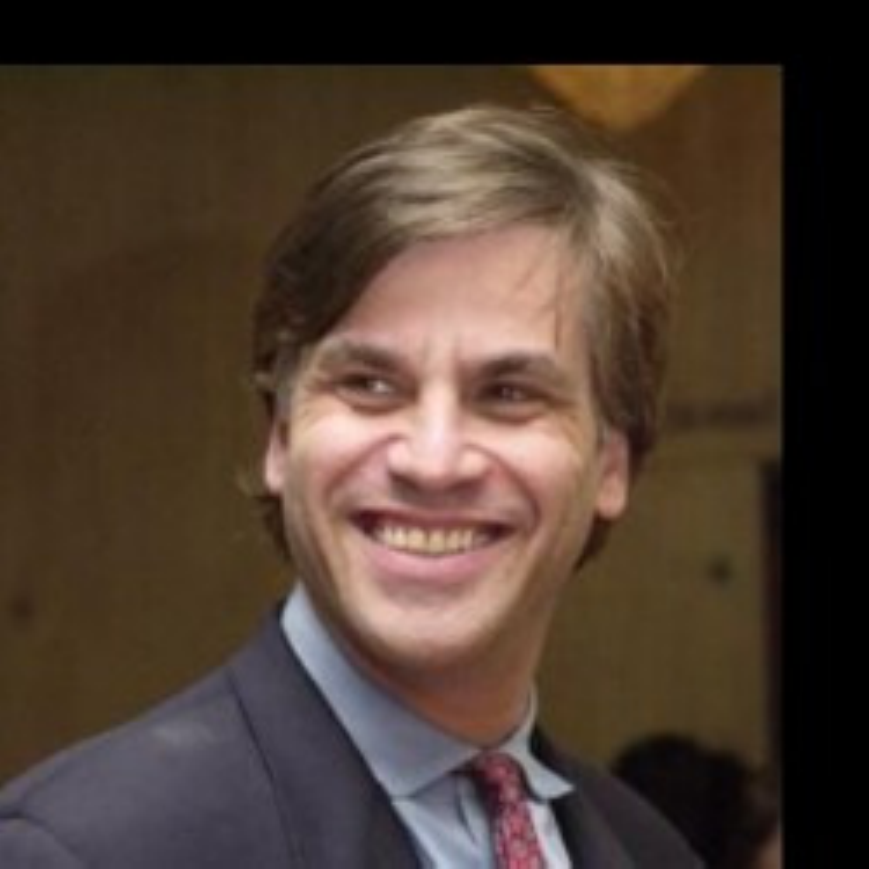}
		\label{fig:entity2}}
	\caption{Two different face images from the same person.}\label{fig:entity}\vspace{-0.15in}
\end{figure}

\vspace{-0.05in}
Since the facial attributes are used for stranger determination, the trained classifier is expected to  make consistent predictions given a specific person. That is, given two different face images of the same person, ideally all the 16 facial attributes obtained from the two images are identical. To evaluate classification consistency, we use the LFW image database \cite{huang2007labeled} that  has been widely used in the literature. Since images in the LFW database are organized by person, it is more efficient to sample images for experiment. In this experiment, we randomly pick 50 persons, and a pair of different face images of each person (see Fig. \ref{fig:entity} as an example). The classification results over the two images in Fig. \ref{fig:entity} are presented in Table \ref{tb:entity}. It can be seen that the classified facial attributes of these two images are exactly the same except `Big Lips', `Brown Hair' and `High Cheekbones'. Out of the 50 persons, the classification results for 32 persons are fully consistent. For the rest 18 persons, 7 persons have 15 identical attributes, 8 persons have 14 identical attributes, and the remaining 3 persons have 13 identical attributes.

\begin{table}[h]
	\footnotesize
	\centering
	\caption{Facial Attributes Classification of Fig. \ref{fig:entity1} and Fig. \ref{fig:entity2}.}
	\begin{tabular}{|c|c|c|}
		\hline
		\textbf{Facial Attributes} & \textbf{Fig. \ref{fig:entity1}} & \textbf{Fig. \ref{fig:entity2}} \\
		\hline
		\textbf{Arched Eyebrows} & No & No \\
		\hline
		\textbf{Bushy Eyebrows} & No & No \\
		\hline
		\textbf{Big Lips} & No & Yes \\
		\hline
		\textbf{Big Nose} & Yes & Yes \\
		\hline
		\textbf{Pointy Nose} & No & No \\
		\hline
		\textbf{Black Hair} & No & No \\
		\hline
		\textbf{Brown Hair} & Yes & No \\
		\hline
		\textbf{Blond Hair} & No & No \\
		\hline
		\textbf{Gray Hair} & No & No \\
		\hline
		\textbf{Eyeglasses} &No  & No \\
		\hline
		\textbf{Bald} &No & No \\
		\hline
		\textbf{High Cheekbones} & No & Yes \\
		\hline
		\textbf{Narrow Eyes} & No & No \\
		\hline
		\textbf{Oval Face} & No & No\\
		\hline
		\textbf{Male} &Yes & Yes\\
		\hline
		\textbf{Young} & No & No \\
		\hline
	\end{tabular}\label{tb:entity}
\end{table}

Besides, we examine the classification consistency on persons with more than 4 face images in the LFW dataset. In particular, we pick 2 pairs of different face images for each of those 10 persons. Then we compare the predicted attributes pair by pair, and hence perform 20-pair comparisons. As Table \ref{tb:consist} shows, 8 pairs of face images are labeled with the exactly the same attributes, and only 6 pairs are labeled with 3 or more different attributes.

\begin{table}[h]
	\footnotesize
	\centering
	\caption{Classification consistency of 10 persons with 2 pairs of face images each.}
	\resizebox{.9\hsize}{!}{
	\begin{tabular}{|c|c|c|c|c|c|c|c|}
		\hline
		\textbf{Number of Identical Attributes} & 16 & 15 & 14 & 13 & 12 & 11 & 10 \\
		\hline
		\textbf{Number of Pairs} & 8 & 4 & 2 & 2 & 2 & 1 & 1 \\
		\hline
	\end{tabular}\label{tb:consist}}
\end{table}

Furthermore, we examine the possibility of two different persons being predicted with identical attributes. We randomly pick 100 persons from the LFW dataset, and perform facial attribute classification on a face image of each person. Then, we compare facial attributes of every person with those of the other 99 persons and hence 4950 pairs are compared in total. Only 144 pairs have exactly the same attributes. All these results show that the classification consistency is high. 

\subsection{Optimal Thresholding}\label{subsec:eval_threshold}\vspace{-0.05in}
The above consistency experiments show that facial attributes of two face images from the same person may not be perfectly identical. Hence, a scheme that depends on exactly matching of facial attributes between two faces will not work for stranger determination. The stranger determination is implemented by thresholding the difference of facial attributes between two compared face images to allow a reasonable difference between these two faces. Hence, it is needed to find a proper threshold. The goal is that we can obtain more true positives without causing too many false positives under the threshold. Here, true positive means two different face images of the same person being determined as the same person. False positive means images of two different persons being determined are from the same person. In this experiment, we pick 50 persons from the LFW database, and two different face images with each person. In order to evaluate false positive, 50 tests are conducted. In each test, we pick one face image from the above 50 persons as the target, and choose another face image from a different person to compare with the target. From the above classification consistency evaluations, we consider 0, 1 and 2 as reasonable threshold candidates and show the results in Table \ref{tb:threeshold}. Based on these results, we choose 1 as the threshold in stranger determination which has good performance in both true positive and false positive. 

\begin{table}[h]
	\footnotesize
	\centering
	\caption{Effectiveness of stranger determination under different threshold with 50 tests.}
	\resizebox{.9\hsize}{!}{
	\begin{tabular}{|c|c|c|c|}
		\hline
		& \textbf{Threshold=0} & \textbf{Threshold=1} & \textbf{Threshold=2} \\
		\hline
		\textbf{\# True Positives} & 36 & 45 & 48 \\
		\hline
		\textbf{\# False Positives} & 1 & 3 & 12\\
		\hline
	\end{tabular}}
	\label{tb:threeshold}
\end{table}\vspace{-0.15in}

\subsection{Effectiveness of Target Filter}\label{subsec:eva_filter}\vspace{-0.05in}
This test aims to examine how well the target filter module can detect the target from a photo. In this experiment, we perform target filter on field photos from two different sources where multiple targets might be in one photo. We use \textit{false filtering rate} to measure the performance, which is defined as the percentage of times when not all targets in the photo are successfully detected or any stranger appearing in the photo is mistakenly detected as the target. 

\begin{figure}
	\centering
	\subfigure[A test photo where the target is successfully detected.]{
		\centering
		\includegraphics[scale=0.59]{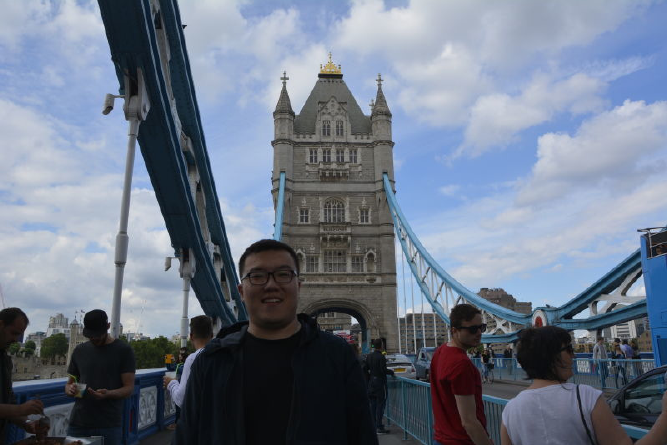}
		\label{fig:target1}}
	\subfigure[A test photo where the target filter failed.]{
		\centering
		\includegraphics[scale=0.09]{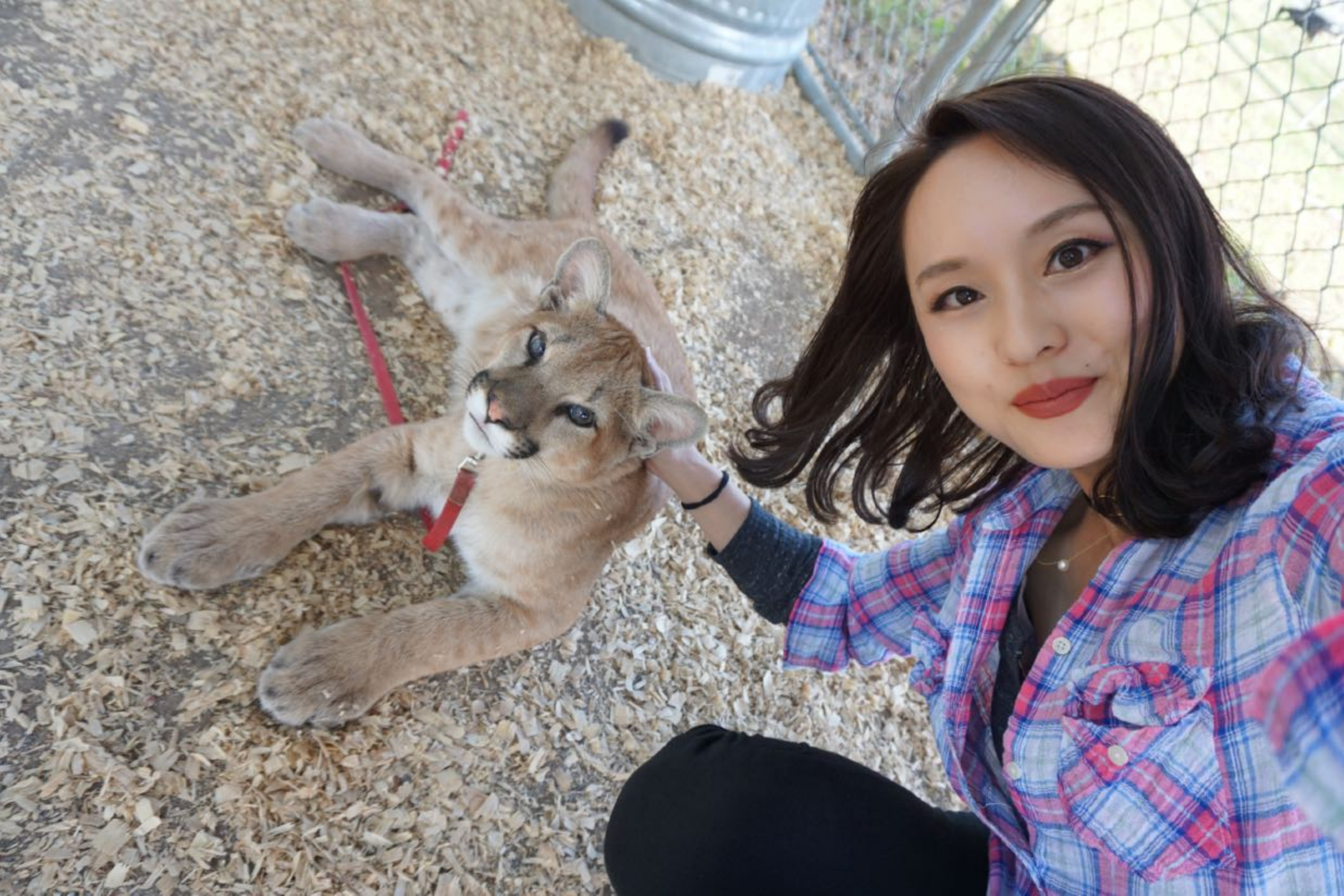}
		\label{fig:target2}}
	\caption{Target filter test on photos taken by the authors.}\label{fig:filter1}\vspace{-0.15in}
\end{figure}

First, we evaluate the effectiveness of target filter on 100 photos, which we have pictured by mobile phone in the past, and at least one target person is included in each photo. The result shows that the false filtering rate is only 8\%, which means the target filter only fails to detect the target in 8 photos. Fig. \ref{fig:filter1} illustrates two example photos of our test. Fig. \ref{fig:target1} is a successful example, but Fig. \ref{fig:target2} is a failed example. The reason for unsuccessful target detection is that the face is not at the central region of the photo, and ‘Smiling’ attribute is falsely predicted as ‘No’. Based on our proposed three rules, only the rule based on face size can be satisfied, and hence the target is not successfully detected.

\begin{figure}
	\centering
	\subfigure{
		\centering
		\includegraphics[scale=0.14]{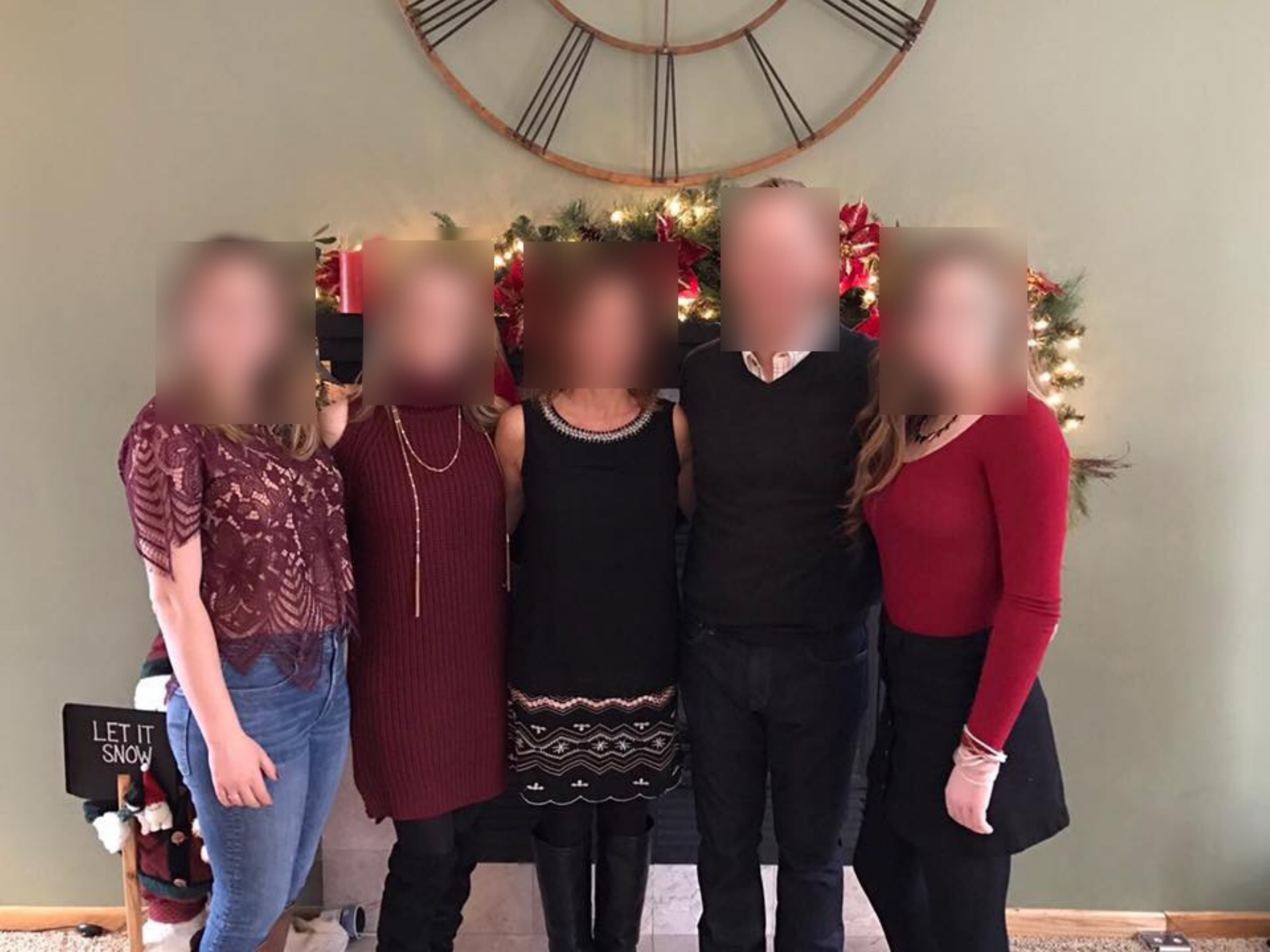}}
	\subfigure{
		\centering
		\includegraphics[scale=0.105]{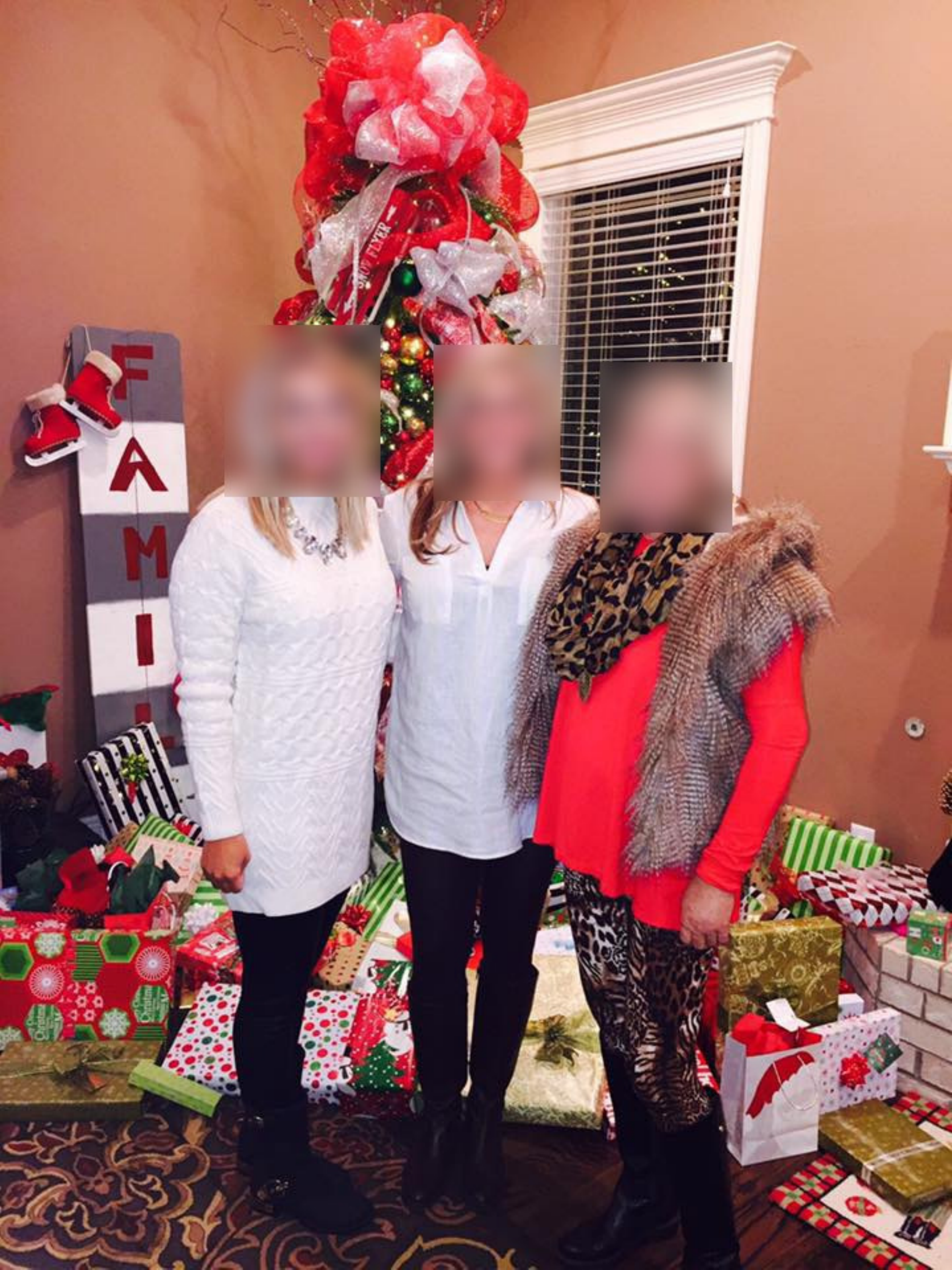}}
	\caption{Target filter test on photos shared by friends on Facebook.}\label{fig:facebook}\vspace{-0.15in}
\end{figure}

Then we pick 100 photos shared by our friends in Facebook from 10/01/2016 to 12/26/2016. At least one target person is included in each photo. Fig. \ref{fig:facebook} shows some example photos, where faces are blurred upon the friends’ request. Similar to the above test, we run target filtering on these 100 photos. The false filtering rate is 12\%, which means the target filtering operation fails in 12 photos. We look into each of those 12 photos, and find the same reason causing false target filtering. When multiple targets are shown in the photo, the target at the rightmost or leftmost is detected as out of the central region of the photo. Also, this target was not smiling when the photo was taken or the ‘Smiling’ attribute is falsely predicted as ‘No’. As a result, in those cases, the rules based on face position and smiling cannot be satisfied, and hence the target filter cannot successfully detect all the targets in the photo. However, the overall target filtering accuracy is still high.

\subsection{Accuracy of Protection}
\begin{figure}[h]
	\footnotesize
	\centering
	\subfigure{
		\includegraphics[scale=0.03,angle=-90]{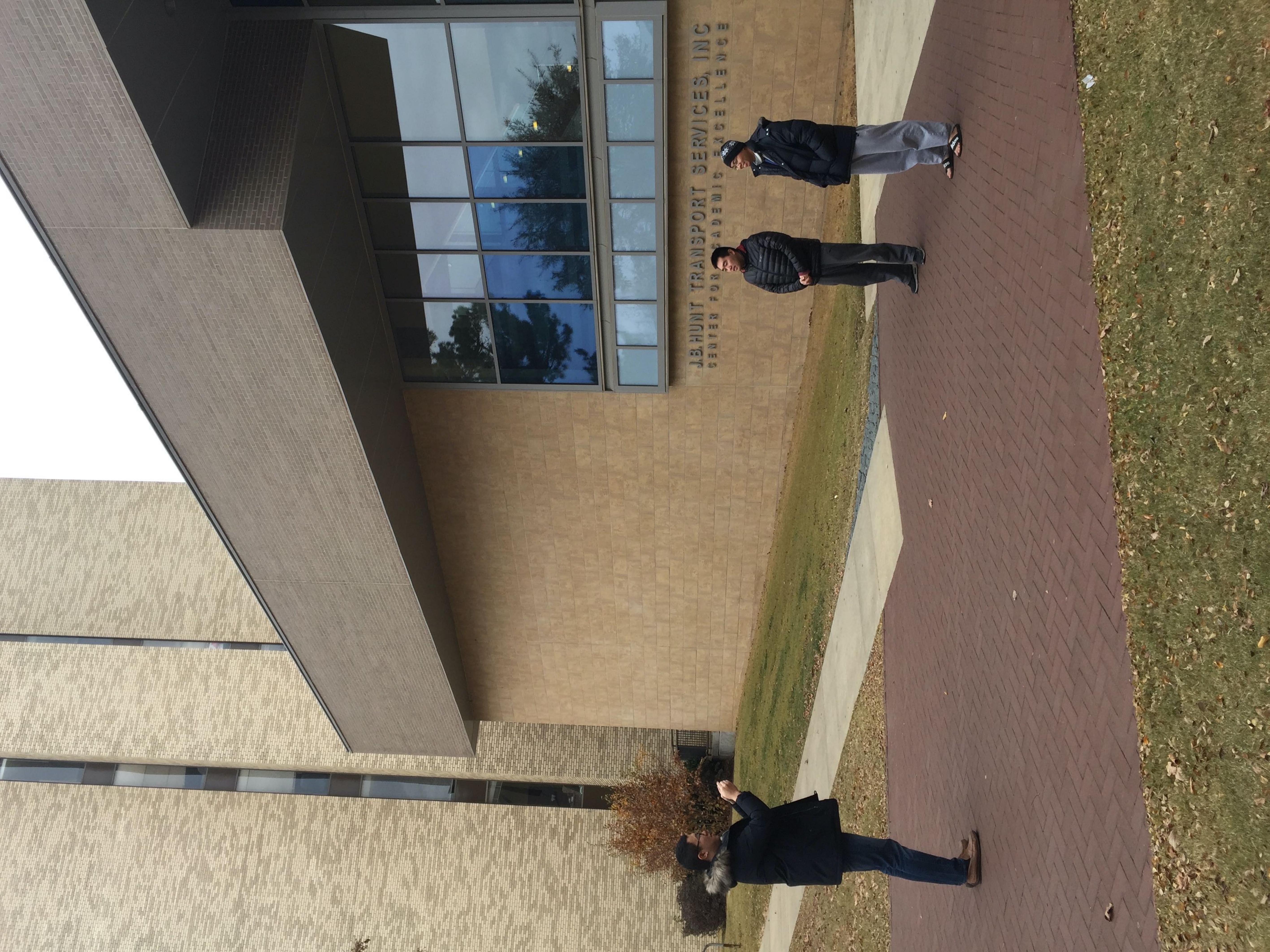}}
	\subfigure{
		\includegraphics[scale=0.03,angle=-90]{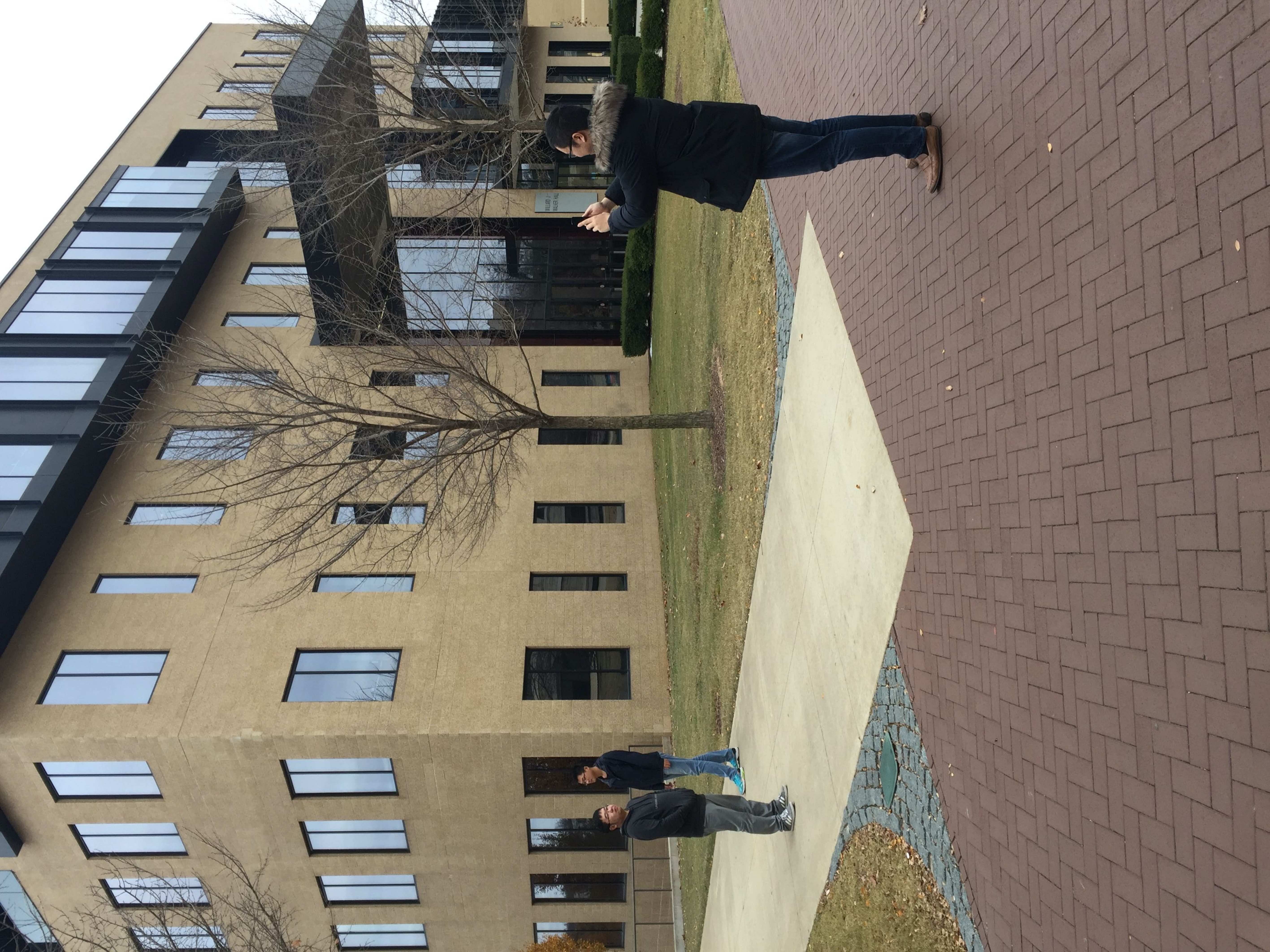}}
	\caption{Example Experiment Scenes}\vspace{-0.15in}
	\label{fig:secne}
\end{figure}

This part evaluates the effectiveness of our system in protecting the stranger's privacy. The experiments are conducted on our campus. Fig. \ref{fig:secne} shows two example experiment scenes. 

\begin{figure}[h]
		\centering
		\includegraphics[scale=0.06]{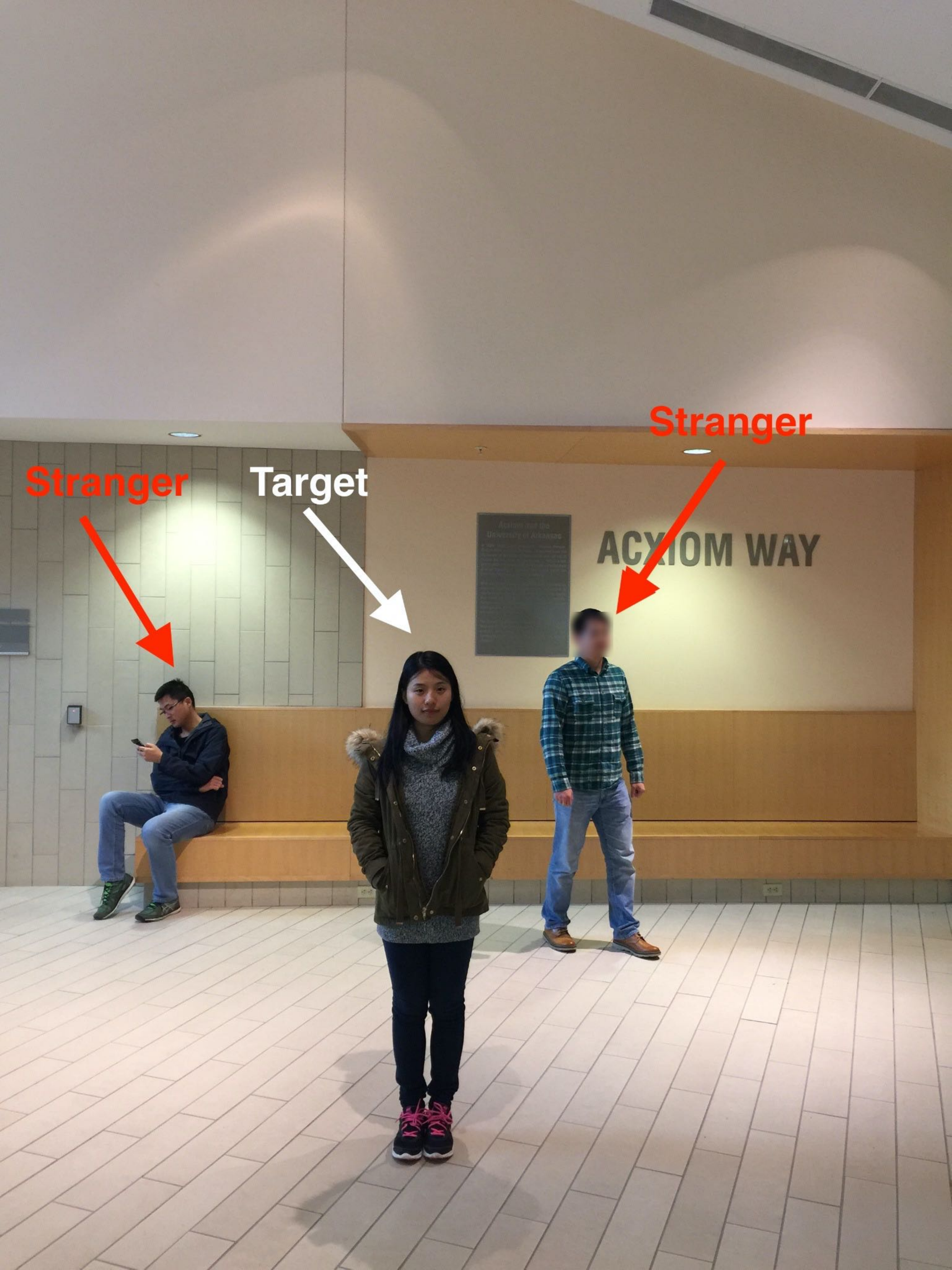}
		\caption{Example of a successful protection.}\vspace{-0.15in}
		\label{fig:blur}
	\end{figure}

\textit{True Protection Rate:} This group of tests considers the scenario where one target person and two strangers appear in the photo. We assume either one of the two strangers or both of them request face blurring. The \textit{true protection rate} is defined as ratio of times when the faces of the requesting strangers are blurred in the photo. For each requesting stranger, we conduct 10 tests separately. Fig. \ref{fig:blur} shows an example where the right stranger's face is successfully blurred. Table \ref{tb:true_protection} shows the true protection rate which is high. 
	
	\begin{table}[h]
		\footnotesize
		\centering
		\caption{True Protection Rate}
		\begin{tabular}{|c|c|}
			\hline
			\# Requesting Strangers & True Protection Rate \\
			\hline
			1 & 90\% \\
			\hline
			2 &  80\%\\
			\hline
		\end{tabular}\label{tb:true_protection}
	\end{table}
	
\textit{False Protection Rate:} Again we consider the scenario where one target person and two strangers appear in the photo. Suppose the two strangers in the photo do not request to blur their faces but other nearby strangers who are not in the photo submit blurring requests. In this case, we define \textit{false protection rate} as the percentage of times when any of two strangers in the photo is mistakenly detected as a requesting stranger and hence falsely blurred. To evaluate the false protection rate in a noisy environment, we conduct simulations with 1, 3, 5 and 10 nearby requesting strangers separately. Specifically, in each test, we randomly pick a certain number of entities from the LFW database, who act as nearby requesting strangers, and one face image for each selected person. For each specific number of requesting strangers, 50 tests are conducted separately. Table \ref{tb:false_protection} shows results with different number of requesting strangers. We can see that false protection rate increases with the increasing number of nearby requesting strangers. This is because the more nearby requesting strangers, the higher possibility of their facial attributes being overlapped with that of strangers in the photo. Note that the false protection rate is as low as 3\% with only one nearby requesting stranger. Even under noisy environment with 3 nearby strangers who request face blurring, the false protection is only 8\%. The false protection rate increases to 24\% with 10 nearby requesting strangers, but this case does not occur often in the real world.
	
	\begin{table}[h]
		\footnotesize
		\centering
		\caption{False Protection Rate}
		\resizebox{.9\hsize}{!}{
		\begin{tabular}{|c|c|c|c|c|}
			\hline
			\multirow{2}{*}{} & \multicolumn{4}{c|}{\textbf{\# Nearby Requesting Strangers}} \\
			\cline{2-5}
			& 1 & 3 & 5 & 10\\
			\hline
			\textbf{False Protection Rate} & 3\% & 8\% & 14\% & 24\%\\
			\hline
		\end{tabular}\label{tb:false_protection}}
	\end{table}\vspace{-0.1in}


\section{Related Work}\label{sec:related}\vspace{-0.05in}
\textbf{Photo and Video Privacy} Jung and Philipose \cite{jung2014courteous} propose a system to protect video privacy. If it detects the person being recorded is making certain gestures like waving hands, the wearable camera will stop recording that person. Raval et al. \cite{raval2014markit} design a system called \textit{MarkIt} to protect video privacy. It detects sensitive objects predefined by users in video, and those sensitive objects will be covered with markers before releasing the video to third-party applications. Jana et al. \cite{jana2013enabling} design an OS abstraction \textit{Recognizer} to enforce fine-grained access control in augmented reality system by reducing the quality of raw sensor data. \textit{Darkly} \cite{jana2013scanner} restricts untrusted applications from accessing raw data from perceptual sensors. 

Schiff et al. \cite{schiff2009respectful} implement a system to protect photo privacy by detecting persons that wear special tracking markers and blurring their faces in photos. However, people who want to protect their privacy must wear special markers beforehand which is not suitable for our considered daily scenarios. Bo et al. \cite{bo2014privacy} propose a protocol to protect the privacy of people being pictured based on a physical tag, which contains their privacy preferences. However, people have to wear clothes with QR-code as privacy tags. Visual fingerprints have also been used to detect whether a user is in a photo \cite{wang2015visually}, but their scheme requires update of visual fingerprints whenever there is any change (e.g., clothes change), requiring too much intervention from people. Templeman et al. \cite{templeman2014reactive} propose an approach to prevent photos from being shared with others by checking the attributes extracted from the photo, such as location and content. \textit{PlaceAvoider} \cite{templeman2014placeavoider} is a context-aware system which can notify the photographer when an application is going to capture photos in sensitive places. Zhang et al. \cite{zhang2016privacy} design a photo capturing and sharing system to protect people's privacy based on graph representations of people's portraits. \textit{Notisense} \cite{pidcock2011notisense} is implemented to notify bystanders of nearby mobile sensing activities. Tan et al. \cite{tan2014short} implement a system to protect photo privacy based on the recognition of persons who are known to the phone owner, and deny third-party applications to access these photos. 

\textbf{Facial Attributes Classification}  Kumar et al. \cite{kumar2008facetracer} propose an approach to train facial attribute classifiers.  Features from manually-picked facial regions for each facial attribute are separately optimized using AdaBoost algorithms. In addition, independent SVM classifiers are trained by feeding optimized features. In this approach, various features are learnt for each facial attribute, and an independent SVM classifier is separately trained. Even though it is a valid approach, it is not efficient for feature extraction and classification. Recently, with the increasing popularity of convolution neural network (CNN), it has been leveraged to extract more sophisticated features of facial attributes. For instance, Kang et al. \cite{kang2015face} propose gated CNNs, which aim to determine which regions of a face are most correlated to corresponding attributes. Then, the output of such CNNs is encoded into a global feature vector for training independent binary SVM classifiers. Zhang et al.  \cite{zhang2015learning}  apply CNNs to learn facial attributes, which are used to infer social relations between pairs of identities with an image. Liu et al. \cite{liu2015deep} design three CNNs, including two localization networks (LNets) and an attribute recognition network (ANet). LNet is designed for localizing features in face images, while ANet is trained on face identities and attributes to extract features. Then, independent SVM classifiers are trained on those extracted features. However, none of them can be directly used for imbalanced distributed datasets.

\section{Conclusion}\label{sec:conclusion}\vspace{-0.05in}
We proposed a system PoliteCamera to protect strangers' privacy who are accidentally captured in a photo taken by mobile phones. The system can inform nearby strangers that they are possibly included in a photo and give them an option to blur their faces in the photo. A novel ABCNN structure is designed to predict facial attributes, where the facial attributes are used to determine whether a requesting stranger is in the photo and which face in the photo belongs to him. We implemented a prototype system, and evaluated its performance through experiments. The accuracy of the facial attributes prediction is better than the state of the art, and experimental evaluations demonstrate that the system can effectively protect strangers’ privacy.\vspace{-0.05in}


	
\bibliographystyle{splncs03}
\bibliography{PERCOM2018}

\end{document}